\documentclass[prb,twocolumn,showpacs,preprintnumbers,amsmath,amssymb]{revtex4}

\usepackage{graphicx}
\usepackage{dcolumn}
\usepackage{bm}

\begin{document}

\title{Domain walls of ferroelectric BaTiO$_3$ within the
Ginzburg-Landau-Devonshire phenomenological model}

\author{P. Marton, I. Rychetsky and J. Hlinka}
\affiliation{Institute of Physics, Academy of Sciences of the Czech
Republic, Na Slovance 2, 18221 Praha 8, Czech Republic}

\date{\today}

\begin{abstract}

Mechanically compatible and electrically neutral domain walls in
tetragonal, orthorhombic and rhombohedral ferroelectric phases of
BaTiO$_3$ are systematically investigated in the framework of the
phenomenological Ginzburg-Landau-Devonshire (GLD) model with
parameters of Ref.\,[Hlinka and Marton, Phys. Rev. {\bf 74},
104104 (2006)]. Polarization and strain profiles within domain
walls are calculated numerically and within an approximation
leading to the quasi-one-dimensional analytic solutions applied
previously to the ferroelectric walls of the tetragonal phase [W.
Cao and L.E. Cross, Phys. Rev. {\bf 44}, 5 (1991)]. Domain wall
thicknesses and energy densities are estimated for all
mechanically compatible and electrically neutral domain wall
species in the entire temperature range of ferroelectric phases.
The model suggests that the lowest energy walls in the
orthorhombic phase of BaTiO$_3$ are the 90-degree and 60-degree
walls. In the rhombohedral phase, the lowest energy walls are the
71-degree and 109-degree walls. All these ferroelastic walls have
thickness below 1\,nm except for the 90-degree wall in the
tetragonal phase and the 60-degree S-wall in the orthorhombic
phase, for which the larger thickness of the order of 5\,nm was
found. The antiparallel walls of the rhombohedral phase have the
largest energy and thus they are unlikely to occur. The
calculation indicates that the lowest energy structure of the
109-degree wall and few other domain walls in the orthorhombic and
rhombohedral phases resemble Bloch-like walls known from
magnetism.

\end{abstract}

\pacs{ 77.80.-e, 77.80.Dj, 77.84.Dy }

\keywords{Domain walls,BaTiO$_3$, Phenomenological Ginnzburg-Landau-Devonshire
model}

\maketitle

\section{INTRODUCTION}

Domain structure is an important ingredient in functionality of ferroelectric
materials. Among others, it has impact on their nonlinear optical properties,
dielectric permittivity and polarization switching phenomena. Since domain
boundaries in ferroelectric perovskite materials can simultaneously play the
role of the ferroelectric and ferroelastic walls, such domain walls also
strongly influence the electromechanical material properties: they facilitate
switching of spontaneous polarization and spontaneous deformation, thus giving
rise to a large extrinsic contribution to e.g. piezoelectric constants, which
makes ferroelectric materials extremely attractive for applications. The domain
structure also provides additional degree of freedom for tuning of material
properties. In general, further development of domain engineering strategies
requires deeper understanding of the physics of ferroelectric domain wall
itself.

The Ginzburg-Landau-Devonshire (GLD) theory provides a feasible
tool for such a purpose. Landau-Devonshire model describes
phase-transition properties of single-domain crystal using a
limited number of parameters, which are determined experimentally
(or recently also using ab-initio methods). Introduction of
Ginzburg gradient term to the free energy functional enables
addressing nonhomogeneous multi-domain ferroelectric state. The
GLD model was previously used for computation of domain wall
properties in ferroelectric materials (e.g.
Refs.\,\onlinecite{art_cao_cross_prb_1991,
art_ishibashi_salje_2002, art_huang_jiang_hu_liu_JPCM_1996,
art_hlinka_marton_prb_2006,
art_erhart_cao_fousek_ferroelectrics_2001}) and in phase-field
computer modeling of domain formation and
evolution.\cite{art_nambu_sagala_1994, art_hu_chen_1998,
art_ahluwalia_2003, art_marton_hlinka_2006} The GLD model can be
regarded as a bridge model covering length-scales inaccessible by
ab-initio and micro-mechanical models.

\begin{figure}
\centerline{
\includegraphics[width=8.6cm, clip=true]{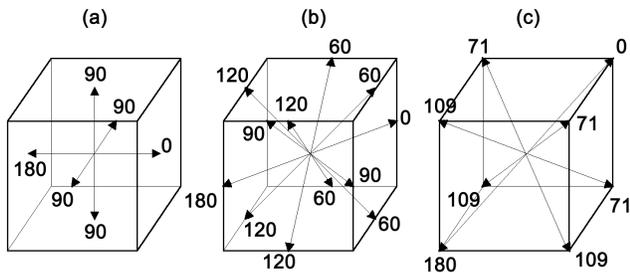}}
\caption{Directions of spontaneous polarization in the (a)
tetragonal, (b) orthorhombic and (c) rhombohedral phase. Angles
between polarization direction '0' and its symmetry equivalent
ones are indicated.}\label{fig_spontanni_stavy}
\end{figure}

BaTiO$_3$ represents a typical ferroelectric material which
undergoes a sequence of phase transitions from high-temperature
paraelectric cubic $Pm3m$ ($O_h$) to the ferroelectric tetragonal
$P4mm$ ($C_{4v}$), orthorhombic $Amm2$ ($C_{2v}$) and rhombohedral
$R3m$ ($C_{3v}$) phase. Energetically equivalent directions of
spontaneous polarization vector, identifying possible
ferroelectric domain states in a particular ferroelectric phase,
are displayed in Fig.\,\ref{fig_spontanni_stavy}. Domain
boundaries separating two domain states are characterized by the
rotational angle needed to match spontaneous polarizations on both
sides of the boundary. For example, the boundary separating
domains with mutually perpendicular spontaneous polarization is
commonly called 90$^\circ$ wall, while the one between
antiparallel spontaneous polarization regions is called
180$^\circ$ wall. Other angles are possible in orthorhombic and
rhombohedral phases of BaTiO$_3$, where the spontaneous
polarization is oriented along the cubic face diagonals and body
diagonals, respectively.

The aim of this paper is to calculate basic characteristics of all
electrically neutral and mechanically compatible domain walls in
all ferroelectric phases of BaTiO$_3$. For a better comparison of
domain wall properties like their thickness or energy density, we
employ an Ising-like approximation leading to previously proposed
analytically solvable one-dimensional
solutions.\cite{art_cao_cross_prb_1991} The paper is organized as
follows. In Section II. we give an overview of the different kinds
of mechanically compatible domain walls in the three ferroelectric
phases. It follows from general theory about macroscopic
mechanical compatibility of adjacent domain
states.\cite{art_fousek_janovec_jap_1969,janovecIT} The GLD
parameters used for calculation of the domain wall properties in
BaTiO$_3$ are the same as in our preceding
work,\cite{art_hlinka_marton_prb_2006} but for the sake of
convenience, the definition and the GLD model and its parameters
are resumed in Section III. Section IV. is devoted to the
description of the computational scheme and approximations applied
here to solve analytically the Euler-Lagrange equations. The main
result of our study - systematic numerical evaluation of
thicknesses, energies, polarization profiles and other properties
for different domain walls, is presented in Section V. Sections
VI. and VII. are devoted to the discussion of validity of used
approximations and the final conclusion, respectively.

\section{\label{sec_domain_walls}MECHANICALLY COMPATIBLE DOMAIN WALLS IN BARIUM TITANATE}

The energy-degeneracy of different directions of spontaneous polarization leads
to the appearance of ferroelectric domain structure. Individual domains are
separated by domain walls, where the polarization changes from one state to
another. Here, only planar domain walls are considered. Orientations of
mechanically compatible domain walls are determined by the equation for
mechanically compatible interfaces separating two domains with the strain
tensors $e_{ij}(-\infty)$ and $e_{ij}(\infty)$ :
\begin{eqnarray}
\sum_{m,n=1}^3[e_{mn}(\infty)-e_{mn}(-\infty)]x_m x_n=0~.
\end{eqnarray}
Systematic analysis of this equation using symmetry arguments has been done e.g.
in Refs.\,\onlinecite{art_fousek_janovec_jap_1969, art_fousek_czp_1971,
janovecIT}. In general, the number $N$ of mechanically compatible domain walls
separating two particular domain states can have only one of the three values:
$N=0$, $N=2$ or $N=\infty$. In case of $N=2$ there exist two mutually
perpendicular domain walls. Each of them is either a crystallographic
($W_f$-type) wall or non-crystallographic ($S$-type) wall. Orientation of the
$W_f$-wall is fixed by symmetry of the crystal, while orientation of the
$S$-wall is determined by components of the strain tensor in adjacent domains
(and its orientation can be therefore dependent on temperature). For $N=\infty$
there exists infinite number of wall orientations, some of them may be preferred
energetically.

\begin{figure}
\centerline{
\includegraphics[width=8.6cm, clip=true]{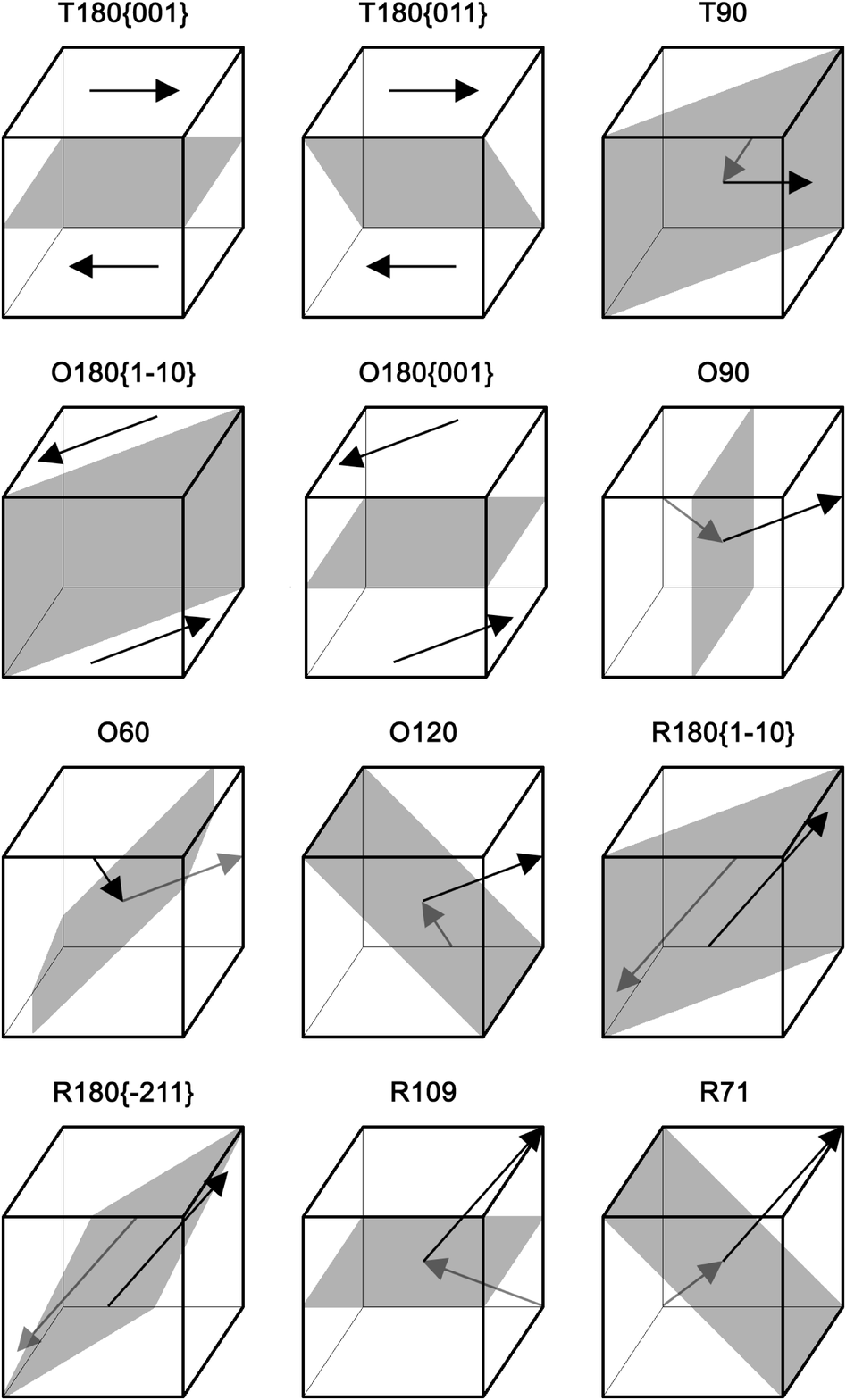}}
\caption{Set of mechanically compatible and electrically neutral
domain walls in the three ferroelectric phases of BaTiO$_3$. In the
case of 180$^\circ$ domain walls, where the orientation is not
determined by symmetry, walls with the most important
crystallographic orientations are displayed.}\label{fig_plochy}
\end{figure}

Further, the electrically neutral domain walls will be
considered.\cite{art_fousek_janovec_jap_1969} It implies that the difference
${\bf P}(\infty)-{\bf P}(-\infty)$ between the spontaneous polarizations in the
adjacent domains is perpendicular to the unit vector ${\bf s}$, normal to the
domain wall: \begin{equation} \label{eq:neutral} \left( {\bf P}(\infty)-{\bf
P}(-\infty) \right)\cdot{\bf s} = 0\,. \end{equation} We also define a unit
vector ${\bf r} \parallel ({\bf P}(\infty)-{\bf P}(-\infty))$, which identifies
the component of the spontaneous polarization which reverses when crossing the
wall. Then the charge neutrality condition (\ref{eq:neutral}) can be expressed
as ${\bf r}\cdot{\bf s} = 0$. Finally, let us introduce a third base vector
${\bf t} = {\bf r}\times {\bf s}$, which complements the symmetry-adapted
orthonormal coordinate system (r, s, t).

BaTiO$_3$ symmetry allows a variety of domain
walls.\cite{art_fousek_czp_1971} Ferroelectric walls of BaTiO$_3$
can be divided in two groups - the
non-ferroelastic\cite{janovecIT} walls separating domains with
antiparallel polarization ($e_{mn}(\infty)-e_{mn}(-\infty)=0$,
$N=\infty$) and the ferroelastic walls with other than 180$^\circ$
between polarization in the adjacent domain states ($N=2$). The
${\bf r}\cdot{\bf s} = 0$ condition implies that the neutral
non-ferroelastic walls are parallel to the spontaneous
polarization, and the neutral ferroelastic walls realize a
"head-to-tail" junction. The set of plausible neutral and
mechanically compatible domain wall types are schematically shown
in Fig.\,\ref{fig_plochy}. Domain walls are labeled by a symbol
composed of the letter specifying the ferroelectric phase (T, O or
R staying for the tetragonal, orthorhombic or rhombohedral,
resp.), number indicating the polarization rotation angle (180,
120, 109, 90, 71 or 60 degrees) and, if needed, the orientation of
the domain wall normal with respect to the parent pseudo-cubic
reference structure.

\begin{table*}
\caption{Cartesian components of switching vectors ${\bf r}$, domain wall
normals ${\bf s}$, and boundary conditions for polarization and strain in
adjacent domain states for the inspected domain walls. Vector ${\bf s}_{\rm
O60}$ is defined in Eqn.\,(\ref{eqn_O60rst_coord}), $P_0$ stands for magnitude
of spontaneous polarization. As usual, spontaneous quantities are those
minimizing GLD functional. Numerical values used in this work are given in
Sec\.\,\ref{sec_results}.}\label{tab_boundary_conditions}
\begin{tabular}{lcccccc}
\hline\hline\
Wall                &{\bf r}                                                        &{\bf s}&${{\bf P}(-\infty)}/{P_0}$                                &${{\bf P}(\infty)}/{P_0}$                                       &${\bf e}(-\infty)$                    &${\bf e}(\infty)$\\
\hline
T180\{001\}         &$(1,0,0)$                                                      &$(0,0,1)$                                                      &$(1,0,0)$                                                      &$(-1,0,0)$                                                         &$(e_\parallel,e_\perp,e_\perp,0,0,0)$  &$(e_\parallel,e_\perp,e_\perp,0,0,0)$\\
T180\{011\}         &$(1,0,0)$                                                      &$(0,\frac{1}{\sqrt{2}},\frac{1}{\sqrt{2}})$                    &$(1,0,0)$                                                      &$(-1,0,0)$                                                         &$(e_\parallel,e_\perp,e_\perp,0,0,0)$  &$(e_\parallel,e_\perp,e_\perp,0,0,0)$\\
T90                 &$(\frac{1}{\sqrt{2}},\frac{1}{\sqrt{2}},0)$            &$(\frac{1}{\sqrt{2}},\frac{-1}{\sqrt{2}},0)$                   &$(1,0,0)$&$(0,-1,0)$&$(e_\parallel,e_\perp,e_\perp,0,0,0)$  &$(e_\perp,e_\parallel,e_\perp,0,0,0)$\\
O180\{1$\bar{1}$0\} &$(\frac{1}{\sqrt{2}},\frac{1}{\sqrt{2}},0)$                    &$(\frac{1}{\sqrt{2}},\frac{-1}{\sqrt{2}},0)$                   &$(\frac{1}{\sqrt{2}},\frac{1}{\sqrt{2}},0)$                    &$(\frac{-1}{\sqrt{2}},\frac{-1}{\sqrt{2}},0)$                      &$(e_{\rm a},e_{\rm a},e_{\rm c},0,0,2e_{\rm b})$               &$(e_{\rm a},e_{\rm a},e_{\rm c},0,0,2e_{\rm b})$\\
O180\{001\}         &$(\frac{1}{\sqrt{2}},\frac{1}{\sqrt{2}},0)$                    &$(0,0,1)$                                                      &$(\frac{1}{\sqrt{2}},\frac{1}{\sqrt{2}},0)$                    &$(\frac{-1}{\sqrt{2}},\frac{-1}{\sqrt{2}},0)$                      &$(e_{\rm a},e_{\rm a},e_{\rm c},0,0,2e_{\rm b})$               &$(e_{\rm a},e_{\rm a},e_{\rm c},0,0,2e_{\rm b})$\\
O90                 &$(0,1,0)$                                                      &$(1,0,0)$                                                      &$(\frac{1}{\sqrt{2}},\frac{1}{\sqrt{2}},0)$                    &$(\frac{1}{\sqrt{2}},\frac{-1}{\sqrt{2}},0)$                       &$(e_{\rm a},e_{\rm a},e_{\rm c},0,0,2e_{\rm b})$               &$(e_{\rm a},e_{\rm a},e_{\rm c},0,0,-2e_{\rm b})$\\
O60                 &$(\frac{1}{\sqrt{2}},0,\frac{1}{\sqrt{2}})$                    &${\bf s}_{\rm O60}$                                            &$(\frac{1}{\sqrt{2}},\frac{1}{\sqrt{2}},0)$                    &$(0,\frac{1}{\sqrt{2}},\frac{-1}{\sqrt{2}})$                       &$(e_{\rm a},e_{\rm a},e_{\rm c},0,0,2e_{\rm b})$               &$(e_{\rm c},e_{\rm a},e_{\rm a},-2e_{\rm b},0,0)$\\
O120                &$(\frac{1}{\sqrt{6}},\frac{2}{\sqrt{6}},\frac{-1}{\sqrt{6}})$  &$(\frac{1}{\sqrt{2}},0,\frac{1}{\sqrt{2}})$                    &$(\frac{1}{\sqrt{2}},\frac{1}{\sqrt{2}},0)$                    &$(0,\frac{-1}{\sqrt{2}},\frac{1}{\sqrt{2}})$                       &$(e_{\rm a},e_{\rm a},e_{\rm c},0,0,2e_{\rm b})$               &$(e_{\rm c},e_{\rm a},e_{\rm a},-2e_{\rm b},0,0)$\\
R180\{1$\bar{1}$0\} &$(\frac{1}{\sqrt{3}},\frac{1}{\sqrt{3}},\frac{1}{\sqrt{3}})$   &$(\frac{1}{\sqrt{2}},\frac{-1}{\sqrt{2}},0)$                   &$(\frac{1}{\sqrt{3}},\frac{1}{\sqrt{3}},\frac{1}{\sqrt{3}})$   &$(\frac{-1}{\sqrt{3}},\frac{-1}{\sqrt{3}},\frac{-1}{\sqrt{3}})$    &$(e_{\rm a},e_{\rm a},e_{\rm a},2e_{\rm b},2e_{\rm b},2e_{\rm b})$         &$(e_{\rm a},e_{\rm a},e_{\rm a},2e_{\rm b},2e_{\rm b},2e_{\rm b})$\\
R180\{$\bar{2}$11\} &$(\frac{1}{\sqrt{3}},\frac{1}{\sqrt{3}},\frac{1}{\sqrt{3}})$   &$(\frac{-2}{\sqrt{6}},\frac{1}{\sqrt{6}},\frac{1}{\sqrt{6}})$  &$(\frac{1}{\sqrt{3}},\frac{1}{\sqrt{3}},\frac{1}{\sqrt{3}})$   &$(\frac{-1}{\sqrt{3}},\frac{-1}{\sqrt{3}},\frac{-1}{\sqrt{3}})$    &$(e_{\rm a},e_{\rm a},e_{\rm a},2e_{\rm b},2e_{\rm b},2e_{\rm b})$         &$(e_{\rm a},e_{\rm a},e_{\rm a},2e_{\rm b},2e_{\rm b},2e_{\rm b})$\\
R109                &$(\frac{1}{\sqrt{2}},\frac{1}{\sqrt{2}},0)$                    &$(0,0,1)$                                                      &$(\frac{1}{\sqrt{3}},\frac{1}{\sqrt{3}},\frac{1}{\sqrt{3}})$   &$(\frac{-1}{\sqrt{3}},\frac{-1}{\sqrt{3}},\frac{1}{\sqrt{3}})$     &$(e_{\rm a},e_{\rm a},e_{\rm a},2e_{\rm b},2e_{\rm b},2e_{\rm b})$         &$(e_{\rm a},e_{\rm a},e_{\rm a},-2e_{\rm b},-2e_{\rm b},2e_{\rm b})$\\
R71                 &$(0,1,0)$                                                      &$(\frac{1}{\sqrt{2}},0,\frac{1}{\sqrt{2}})$                    &$(\frac{1}{\sqrt{3}},\frac{1}{\sqrt{3}},\frac{1}{\sqrt{3}})$   &$(\frac{1}{\sqrt{3}},\frac{-1}{\sqrt{3}},\frac{1}{\sqrt{3}})$      &$(e_{\rm a},e_{\rm a},e_{\rm a},2e_{\rm b},2e_{\rm b},2e_{\rm b})$         &$(e_{\rm a},e_{\rm a},e_{\rm a},-2e_{\rm b},2e_{\rm b},-2e_{\rm b})$\\
\hline\hline
\end{tabular}
\end{table*}

Our choice of the base vectors ${\bf r}, {\bf s}$ and of the
spontaneous polarization and strain components in the adjacent
domain pairs for each domain wall type shown in
Fig.\,\ref{fig_plochy} are summarized in
Table\,\ref{tab_boundary_conditions}. Base vectors coincide with
special crystallographical directions, except for the O60 wall
where the ${\bf s}$ and ${\bf t}$ vectors depend on the
orthorhombic spontaneous strain (see
Table\,\ref{tab_boundary_conditions}) as
follows:\cite{art_erhart_cao_fousek_ferroelectrics_2001}
\begin{eqnarray}
{\bf r}_{\rm O60}&=&\left(\frac{1}{\sqrt{2}},0,\frac{1}{\sqrt{2}}\right)\nonumber\\
{\bf s}_{\rm O60}&=&\left(\frac{e_{\rm a}-e_{\rm c}}{D_1},\frac{2e_{\rm b}}{D_1},\frac{e_{\rm c}-e_{\rm a}}{D_1}\right)\nonumber\\
{\bf t}_{\rm O60}&=&\left(\frac{-e_{\rm b}}{D_2},\frac{e_{\rm
a}-e_{\rm c}}{D_2},\frac{e_{\rm
b}}{D_2}\right)~~,\label{eqn_O60rst_coord}
\end{eqnarray}
with $D_1=\sqrt{2}D_2$, $D_2=\sqrt{(e_{\rm a}-e_{\rm c})^2+2e_{\rm
b}^2}$ and with $e_{\rm a}, e_{\rm b}$, and  $e_{\rm c}$  defined in
Table\,I.

Although only the neutral walls are discussed in the following, the
Fig.\,\ref{fig_plochy} is actually helpful in enumeration of all possible
mechanically compatible domain wall species in BaTiO$_3$. In principle,
mechanical compatibility allows 180$^\circ$ $W_\infty$-type domain walls with an
arbitrary orientation of the domain wall in all three ferroelectric phases
(T180, O180, R180). Obviously, they are electrically neutral only if the domain
wall normal is parallel with the spontaneous polarization. Ferroelastic walls
exists in mutually perpendicular pairs. In the tetragonal phase, there exist
90$^\circ$ $W_f$-type domain walls (T90), either charged (head-to-head or
tail-to-tail) or neutral (head-to-tail). The orthorhombic phase is more complex.
In the case of 60$^\circ$ angle between polarization directions, the N=2 pair is
formed by a charged $W_f$-type wall and neutral $S$-type wall. The case of
120$^\circ$ angle is similar but $W_f$-wall is neutral and $S$-wall is charged.
In addition, there are again charged or neutral 90$^\circ$ $W_f$-walls (O90).
The rhombohedral phase has pairs of charged and neutral $W_f$-type domain walls
with the angle between polarizations either 109$^\circ$ or 71$^\circ$ (R109 or
R71, resp.). Since only neutral walls are discussed here, the $S$-type domain
wall will be referred to as O60 and $W_f$ wall as O120.

\section{\label{sec_energy-expansion}GLD MODEL FOR BARIUM TITANATE}

Calculations presented in this paper are based on the GLD model with
anisotropic gradient terms,  reviewed in
Ref.\,\onlinecite{art_hlinka_marton_prb_2006}. The free energy $F$
is expressed in terms of polarization and strain field taken for
primary and secondary order-parameter, resp.:
\begin{eqnarray}\label{eqn_total_potential}
F\left[\{ P_i, P_{i,j}, e_{ij} \}\right]=\int f {\rm d{\bf r}}~,
\end{eqnarray}
where the free energy density $f$ consists of Landau,
gradient, elastic and electrostriction part
\begin{eqnarray}
f=f_{\rm L}^{(e)}\{P_i\}+f_{\rm C}\{P_i,e_{ij}\}+f_{\rm
q}\{P_i,e_{ij}\}+f_{\rm
G}\{P_{i,j}\}.
\end{eqnarray}
The Landau potential considered here is expanded up to the sixth
order in components of polarization for the cubic symmetry
($O_h^1$):
\begin{eqnarray}
f_{\rm L}^{\rm (e)}=&&\alpha_{1}\left(P_1^2+P_2^2+P_3^2\right) \nonumber\\
            &&+ \alpha_{11}^{\rm (e)}\left(P_1^4+P_2^4+P_3^4\right) \nonumber\\
            &&+ \alpha_{12}^{\rm (e)}\left(P_1^2P_2^2+P_2^2P_3^2+P_1^2P_3^2\right) \nonumber\\
            &&+ \alpha_{111}\left(P_1^6+P_2^6+P_3^6\right) \nonumber\\
            &&+ \alpha_{112}(P_1^4(P_2^2+P_3^2)+P_2^4(P_1^2+P_3^2)\nonumber\\
            &&\,\,\,\,\,+ P_3^4(P_1^2+P_2^2)).\nonumber\\
            &&+ \alpha_{123}P_1^2P_2^2P_3^2\label{eqn_landauexpansion}
\end{eqnarray}
with the three temperature-dependent coefficients
$\alpha_{1},\alpha_{11}$ and $\alpha_{111}$, as in
Ref.\,\onlinecite{art_bell_jap_2000}. This expansion produces the
6 equivalent domain states in the tetragonal phase, 12 in the
orthorhombic, and 8 in the rhombohedral phase (see
Fig.\,\ref{fig_spontanni_stavy}).

Dependence of the free energy on the strain is encountered by including
elastic and linear-quadratic electrostriction functionals $F_{\rm C}$ and
$F_{\rm q}$, resp. Their corresponding free energy densities are
\begin{eqnarray}
f_{\rm C}=\frac{1}{2}e_{\rho}C_{\rho
\sigma}e_{\sigma}\label{eqn_elastic_energy}
\end{eqnarray}
and
\begin{eqnarray}
f_{\rm q}=-q_{ijkl}e_{ij}P_k P_l~,\label{eqn_electrostriction_energy}
\end{eqnarray}
where $e_{ij}=\frac{1}{2}\left(\partial u_i/\partial x_j+\partial u_j/\partial
x_i\right)$ and $C_{\alpha\beta}$, $q_{\alpha\beta}$ are components of
elastic and electrostriction tensor in Voigt notation, $C_{11}=C_{1111}$,
$C_{12}=C_{1122}$, $C_{44}=C_{1212}$, $q_{11}=q_{1111}$, $q_{12}=q_{1122}$, but
$q_{44}=2q_{1122}$.

The elastic and electrostriction terms result in re-normalization of
the bar expansion coefficients $\alpha_{11}^{(e)}$ and
$\alpha_{12}^{(e)}$ when minimizing the free energy with respect to
strains (in the homogeneous sample). The bar $\alpha_{11}^{(e)},\
\alpha_{12}^{(e)}$ and the relaxed $\alpha_{11},\ \alpha_{12}$
coefficients are related as:\cite{art_hlinka_marton_prb_2006}

\begin{eqnarray}
\alpha_{11}^{(e)}&=&\alpha_{11}+\frac{1}{6}\left[\frac{\hat{q}_{11}^2}{\hat{C}_{11}}+2\frac{\hat{q}_{22}^2}{\hat{C}_{22}}\right]\nonumber\\
\alpha_{12}^{(e)}&=&\alpha_{12}+\frac{1}{6}\left[2\frac{\hat{q}_{11}^2}{\hat{C}_{11}}-2\frac{\hat{q}_{22}^2}{\hat{C}_{22}}+3\frac{q_{44}^2}{C_{44}}\right]\label{eqn_alpha_clamped}
\end{eqnarray}
with
\begin{eqnarray}
\hat{C}_{11}&=&C_{11}+2C_{12}\nonumber\\
\hat{C}_{12}&=&C_{11}-C_{12}\nonumber\\
\hat{q}_{11}&=&q_{11}+2q_{12}\nonumber\\
\hat{q}_{12}&=&q_{11}-q_{12}~.
\end{eqnarray}

The Ginzburg gradient term $f_{\rm G}$ is considered in the form
\begin{eqnarray}
f_{\rm G}&=&\frac{1}{2}G_{11}(P_{1,1}^2+P_{2,2}^2+P_{3,3}^2)\nonumber\\
&&+ G_{12}(P_{1,1}P_{2,2}+P_{2,2}P_{3,3}+P_{1,1}P_{3,3})\nonumber\\
&&+
\frac{1}{2}G_{44}(\left(P_{1,2}+P_{2,1}\right)^2+\left(P_{2,3}+P_{3,2}\right)^2\nonumber\\
&&\,\,\,\,\,+\left(P_{3,1}+P_{1,3}\right)^2)~.\label{eqn_gradient-direct}
\end{eqnarray}
It was pointed out\cite{art_hlinka_marton_prb_2006} that the tensor of gradient
constants of BaTiO$_3$ is highly anisotropic with fundamental consequences
 on predicted domain wall
properties. Up to now, the isotropic gradient tensor $G_{i,j}$ was mostly
 employed in the computations.

The material-specific coefficients in the model are assumed being constant,
except for the three Landau potential coefficients
\begin{eqnarray}
\alpha_{1}&=&3.34\times10^5(T-381)\nonumber\\
\alpha_{11}&=&4.69\times10^6(T-393)-2.02\times10^8\nonumber\\
\alpha_{111}&=&-5.52\times10^7(T-393)+2.76\times10^9~,
\end{eqnarray}
where $T$ is absolute temperature.\cite{art_bell_jap_2000} The phase transitions
occur in this model at the temperatures 392.3\,K (C$\rightarrow$T), 282.5\,K
(T$\rightarrow$O), and 201.8\,K (O$\rightarrow$R).\cite{art_bell_jap_2000} All
phase transitions are of the first order, the local minima corresponding to the
tetragonal, orthorhombic and rhombohedral phase exist for this Landau potential
between 237\,K and 393\,K, between 104\,K and 303\,K, and below 256\,K,
respectively. Full set of temperature independent parameters of the GLD model
reads:\cite{art_hlinka_marton_prb_2006,note_q44}
$\alpha_{12}=3.230\times\rm10^8\,J m^5 C^{-4}$,
$\alpha_{112}=4.470\times\rm10^9\,J m^{9}C^{-6}$,
$\alpha_{123}=4.910\times\rm10^9\,J m^{9}C^{-6}$,
$G_{11}=51\times\rm10^{-11}\,J m^3 C^{-2}$,
$G_{12}=-2\times\rm10^{-11}\,Jm^3C^{-2}$,
$G_{44}=2\times\rm10^{-11}\,J m^3 C^{-2}$,
$q_{11}=14.20\times\rm10^9\,J m  C^{-2}$,
$q_{12}=-0.74\times\rm10^9\,J m  C^{-2}$,
$q_{44}=1.57\times\rm10^9\,J m C^{-2}$,
$C_{11}=27.50\times\rm10^{10}\,J m^{-3}$,
$C_{12}=17.90\times\rm10^{10}\,J m^{-3}$ and
$C_{44}=5.43\times\rm10^{10}\,J m^{-3}$.

\section{\label{sec_one-dimensional-model} STRAIGHT POLARIZATION PATH APPROXIMATION}

Let us consider a single mechanically compatible and electrically
neutral domain wall in a perfect infinite stress-free crystal.
Within the continuum GLD theory, such domain wall is associated
with a planar kink solution of the Euler-Lagrange equations of the
GLD functional for polarization vector and the strain
tensor.\cite{art_zhirnov_1958, art_ishibashi_salje_2002,
art_cao_cross_prb_1991, art_hlinka_marton_prb_2006} Domain wall
type is specified by selection of the wall normal $\bf s$ and by
the two domain states at $s = -\infty$ and $+\infty$. This ideal
geometry implies that the polarization and strain vary only along
normal to the wall ${\bf s}$ and domain wall can be thus
considered as a trajectory in order-parameter space.

Even if the domain wall is neutral, in its central part the local electric
charges can occur due to the position-dependent polarization. The additional
assumption $\nabla \cdot {\bf {\rm P}}=0$ ensures the absence of charges in the
whole domain wall. Then the local electric field is zero and the electrostatic
contribution vanishes. Such strictly charge-free
solutions\cite{art_zhirnov_1958} were found to be an excellent approximation for
ideal dielectric materials.\cite{art_hlinka_marton_prb_2006} The condition
$\nabla \cdot {\bf {\rm P}}=0$ implies that the polarization vector variation is
restricted to a plane perpendicular to $\bf s$ (the trajectory is constrained to
$P_{\rm s}=P_{\rm s}(\pm\infty)$ plane).

In fact, the polarization trajectories representing the T180 and T90 walls
calculated under $\nabla \cdot {\bf {\rm P}}=0$ constraint were found to be the
straight lines connecting the boundary values.\cite{art_zhirnov_1958,
art_ishibashi_salje_2002, art_cao_cross_prb_1991, art_hlinka_marton_prb_2006}
This greatly simplifies the algebra and the solution of the variational problem
can be found analytically. Therefore, we decided to impose the condition of a
direct, straight polarization trajectory for the variational problem of all
domain wall species of BaTiO$_3$. This condition, further referred as straight
polarization path (SPP) approximation, implies that both $P_{\rm s}$ and $P_{\rm
t}$ polarization components are constant across the wall.  We shall come back to
the meaning and possible drawbacks of this approximation in Section VI.

Polarization and strain in the mechanically compatible and
electrically neutral SPP walls can be cast in the form: ${\bf
P}=[P_{\rm r}(s), P_{\rm s}(\pm\infty), P_{\rm t}(\pm\infty)]$ and
$e=[e_{\rm rr}(\pm\infty), e_{\rm ss}(s), e_{\rm tt}(\pm\infty),
2e_{\rm st}(s), 2e_{\rm rt}(\pm\infty), 2e_{\rm rs}(s)]$,
respectively. The $s$-dependent strain components are calculated
from the mechanical equilibrium condition
\begin{eqnarray}
\sum_{j=1}^3\frac{\partial \sigma_{ij}}{\partial
x_j}=\sum_{j=1}^3\frac{\partial}{\partial x_j}\left(\frac{\partial
f}{\partial e_{ij}}\right)=0~.\label{eqn_mechanical-equilibrium}
\end{eqnarray}
 Boundary conditions for stress and the
fact that all quantities vary only along direction $s$ imply
\begin{eqnarray}
\frac{\partial f_{\rm Cq}}{\partial
e_{ij}}+C=0\label{eqn_mechanical-equilibrium-constant}
\end{eqnarray}
for $ij\in\{\rm ss,st,rs\}$ with the integration constant $C$ to be determined from
boundary conditions.

The Euler-Lagrange equation for polarization reduces to
\begin{eqnarray}
\frac{\partial}{\partial s}\frac{\partial f}{\partial P_{{\rm
r},{\rm s}}}-\frac{\partial f}{\partial P_{\rm r}}=0~.
\end{eqnarray}
where the strain components from
Eqn.\,(\ref{eqn_mechanical-equilibrium-constant}) were substituted
into $f$. Thus, the elastic field was eliminated. It turns out that
the resulting Euler-Lagrange equation for $P_{\rm r}$ is possible to
rewrite in the form
\begin{eqnarray}
\label{eq:el1d}
g\frac{d^2p(s)}{ds^2}=2a_1p(s)+4a_{11}p^3(s)+6a_{111}p^5(s)~,\label{eqn_general_wall}
\end{eqnarray}
where $p(s)$ stands for $P_{\rm r}(s)$, and where the coefficients
$g, a_1, a_{11}$ and $a_{111}$, different for each domain type,
depend only on the material tensors. The boundary values are
$-P_{\rm r}(\infty)=P_{\rm r}(-\infty)=p_{\infty}$.

\begin{figure}
\centerline{
\includegraphics[width=8.6cm, clip=true]{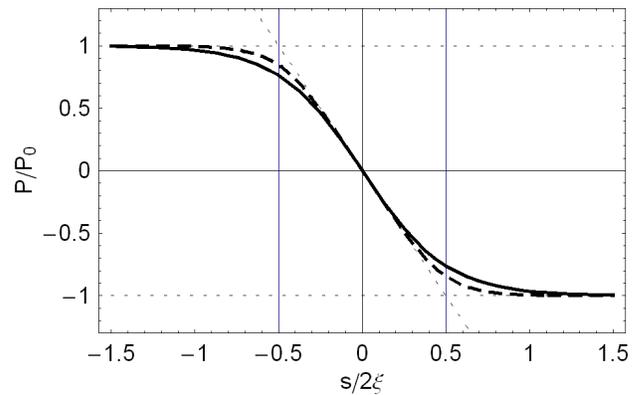}}
\caption{ Profile of the "reversed" polarization component $P_{\rm r
}$ (Eqn.\,\ref{eqn_wall_profile}) for a domain wall with the shape
factor $A=1$ (solid) and $A=3$ (dashed). Both profiles have the same
derivative for $s/2\xi=0$ (dotted) and therefore also the same thickness
(indicated by vertical lines) according to the definition in
Eqn.\,(\ref{eqn_analytical-wall-thickness}).
}\label{fig_clanek_profile}
\end{figure}

Solution of the Euler-Lagrange equation (\ref{eq:el1d}) is well
known.\cite{art_zhirnov_1958, art_houchmandzadeh_lajzerowic_1991,
art_hudak, art_cao_cross_prb_1991, art_ishibashi_dvorak_1976} We
shall follow the procedure of
Ref.\,\onlinecite{art_hlinka_marton_intgrferro}. Integrating Eqn.
(\ref{eq:el1d}) one can obtain the equation:
\begin{eqnarray}
\frac{g}{2}\left(\frac{\partial p}{\partial s}\right)^2=f_{\rm EL}(p)
\end{eqnarray}
where (see Ref.\,\onlinecite{art_hlinka_marton_prb_2006})
\begin{eqnarray}
f_{EL}(p)=a_1 p(s)^2+a_{11} p(s)^4+a_{111} p(s)^6~.
\end{eqnarray}
The function $f_{\rm EL}(p)$ is a double-well "Euler-Lagrange"
potential with two minima $\pm p_{\infty}$, where
\begin{eqnarray}
p_{\infty}^2=\frac{-a_{11}+\sqrt{a_{11}^2-3a_{111}a_{1}}}{3a_{111}}.\label{eqn_19}
\end{eqnarray}
The differential Eqn.\,(\ref{eqn_general_wall}) has the analytical solution
\begin{eqnarray}
p(s)=p_{\infty}\frac{{\sinh}({s}/{\xi^\prime})}{\sqrt{A+{\sinh}^2({s}/{\xi^\prime})}}~,\label{eqn_wall_profile}
\end{eqnarray}
where
\begin{eqnarray}
A=\frac{3a_{111}p_\infty^2+a_{11}}{2a_{111}p_\infty^2+a_{11}}~.
\end{eqnarray}
and
\begin{eqnarray}
\xi^\prime=\frac{\xi}{\sqrt{A}}~.
\end{eqnarray}
Quantity $A$ determines deviation of the profile (\ref{eqn_wall_profile})
from the $tanh$ profile, which occurs
for the $4^{th}$-order potential (i.e., $a_{111}=0,\ A=1$).

The domain wall thickness (Fig.\,\ref{fig_clanek_profile}) is defined as
\begin{eqnarray}
2\xi=p_\infty\sqrt{\frac{2g}{U}}~.\label{eqn_analytical-wall-thickness}
\end{eqnarray}
with $U$ being the energy barrier between the domain states:
\begin{eqnarray}
U=f_{EL}(0)-f_{EL}(p_\infty)=2a_{111}p_\infty^6+a_{11}p_\infty^4~.\label{eqn_energy_on_pinfty}
\end{eqnarray}
The surface energy density of the domain wall is
\begin{eqnarray}
\Sigma&=&\int_{-\infty}^{\infty}\left(f_{EL}(s)-f_{EL}(p_\infty)\right)ds\nonumber\\
&=&\frac{4}{3}p_\infty \sqrt{2gU}\left[A^{5/2}I(A)\right]~,\label{correction_factor_sigma}
\end{eqnarray}
where
\begin{eqnarray}
I(A)=\frac{3}{4}\int_{-\infty}^{\infty}\frac{{\rm
cosh}^2(h)dh}{(A+{\rm cosh}^2(h)-1)^3}~.
\end{eqnarray}

The domain wall characteristics depend on the coefficients $g,\
a_1,\ a_{11},\ a_{111}$ through Eqs. (\ref{eqn_19},
\ref{eqn_wall_profile}, \ref{eqn_analytical-wall-thickness},
\ref{correction_factor_sigma}). The explicit expressions for these
coefficients are summarized for various domain walls in Table\,II.
The expressions are simplified using the notation inspired by
Ref.\,\onlinecite{art_cao_cross_prb_1991}. For all phases we are
using:
\begin{eqnarray}
a_{1}^{\rm
r}&=&\alpha_{1}-\left[\frac{1}{3}\frac{\hat{q}_{11}^2}{\hat{C}_{11}}+\frac{1}{6}\frac{\hat{q}_{22}^2}{\hat{C}_{22}}-\frac{(q_{11}+q_{12})q_{12}^\prime}{2C_{11}^\prime}\right]P_0^2\nonumber
\end{eqnarray}
\begin{eqnarray}
a_{11}^{\rm
r}&=&\frac{\alpha_{11}^{(e)}}{2}+\frac{\alpha_{12}^{(e)}}{4}-\frac{q_{12}^{\prime
2}}{2C_{11}^\prime}\nonumber\\
a_{12}^{\rm
rs}&=&3\alpha_{11}^{(e)}-\frac{\alpha_{12}^{(e)}}{2}-\frac{q_{11}^\prime
q_{12}^\prime}{C_{11}^\prime}-\frac{\hat{q}_{22}^2}{2\hat{C}_{22}}\nonumber
\end{eqnarray}
\begin{eqnarray}
a_{111}^\prime&=&\frac{1}{4}\left(\alpha_{111}+\alpha_{112}\right)\nonumber\\
a_{112}^\prime&=&\frac{1}{4}\left(15\alpha_{111}-\alpha_{112}\right)~,\nonumber\label{eqn_t90elcoefficients}
\end{eqnarray}
for tetragonal and orthorhombic phases we are abbreviating
\begin{eqnarray}
C_{11}^\prime&=&\frac{C_{11}+C_{12}+2C_{44}}{2}\nonumber\\
C_{12}^\prime&=&\frac{C_{11}+C_{12}-2C_{44}}{2}\nonumber\\
C_{66}^\prime&=&\frac{C_{11}-C_{12}}{2}\nonumber
\end{eqnarray}
\begin{eqnarray}
q_{11}^\prime&=&\frac{q_{11}+q_{12}+q_{44}}{2}\nonumber\\
q_{12}^\prime&=&\frac{q_{11}+q_{12}-q_{44}}{2}\nonumber\\
q_{66}^\prime&=&q_{11}-q_{12}~,\nonumber\nonumber\label{eqn_transformace45}
\end{eqnarray}
and for the rhombohedral phase
\begin{eqnarray}
C_{33}^\prime&=&\frac{C_{11}+C_{12}+2C_{44}}{2}\nonumber\\
q_{11}^\prime&=&\frac{q_{11}+2q_{12}+2q_{44}}{3}\nonumber\\
q_{13}^\prime&=&\frac{q_{11}+2q_{12}-q_{44}}{3}~.\label{eqn_transformation_rhombo}
\end{eqnarray}

The expressions for the coefficients of the T180 and T90 domain
walls are equivalent to the previously published
expressions.\cite{art_cao_cross_prb_1991,
art_hlinka_marton_prb_2006} Derivations for O60, O120, and
R180\{$\bar{2}$11\} walls lead to complicated formulas, and
therefore only numerical results for $g, a_1, a_{11}$, and $a_{111}$
coefficients are presented here.

\begin{table*}
\caption{ Parameters characterizing various mechanically compatible
and neutral domain wall species of BaTiO$_3$-like ferroelectrics
within the SPP treatment described in the Section IV. Results for
O60, O120, and R180\{$\bar{2}$11\} walls as well as a few other
parameters are omitted because the corresponding analytical
expressions through the GLD model parameters and spontaneous values
of polarization and strain are too complicated.
}\label{tab_analytical_expressions}
\begin{tabular}{lccccc}
\hline\hline
Domain wall&$p_\infty$&$g$&$a_{1}$&$a_{11}$&$a_{111}$\\
\hline
\multicolumn{6}{c}{Tetragonal phase}\\
T180\{001\}                         &$P_0$                  &$G_{44}$                           &$\alpha_1-e_\parallel q_{11}-e_\perp q_{12}+\frac{C_{12}}{C_{11}}(e_\parallel+e_\perp)q_{12}$                                                                                                &$\alpha_{11}^{(e)}-\frac{1}{2}\frac{q_{12}^2}{C_{11}}$                                                 &$\alpha_{111}$\\
T180\{011\}                         &$P_0$                  &$G_{44}$                           &$\alpha_1-e_\parallel q_{11}+\frac{C_{12}}{C_{11}^\prime}e_\parallel q_{12}-2\frac{C_{44}}{C_{11}^\prime}e_\perp q_{12}$                                                                                      &$\alpha_{11}^{(e)}-\frac{1}{2}\frac{q_{12}^2}{C_{11}^\prime}$                                          &$\alpha_{111}$\\
T90                                 &$\frac{P_0}{\sqrt{2}}$ &$\frac{G_{11}-G_{12}}{2}$        &$\alpha_1^{\rm r}+\alpha_{12}^{\rm rs}\frac{P_0^2}{2}+\alpha_{112}^\prime\frac{P_0^4}{4}$                                                                                            &$\alpha_{11}^{\rm r}+\alpha_{112}^\prime\frac{P_0^2}{2}$                                                     &$\alpha_{111}^\prime$\\
\\
\hline
\multicolumn{6}{c}{Orthorhombic phase}\\
O180\{1$\bar{1}$0\}                 &$P_0$                  &$\frac{G_{11}-G_{12}}{2}$          &$\alpha_1-e_{\rm b}(q_{11}^\prime-q_{12}^\prime)-e_{\rm c} q_{12}-e_{\rm a}(q_{11}^\prime+q_{12}^\prime)$                                                                  &$\frac{\alpha_{11}^{(e)}}{2}+\frac{\alpha_{12}^{(e)}}{4}-\frac{q_{12}^{\prime 2}}{2 C_{11}^\prime}$    &$\frac{1}{4}\left(\alpha_{111}+\alpha_{112}\right)$\\
                                    &$ $                    &$ $                                &$+\frac{q_{12}^{\prime 2}}{C_{11}^\prime}P_0^2$                                                                                                                            &$ $                                                                                                    &$ $\\
O180\{001\}                         &$P_0$                  &$G_{44}$                           &$\alpha_1-e_{\rm a}(q_{11}+q_{12})-e_{\rm b} q_{44}+2e_{\rm a} q_{12}\frac{C_{12}}{C_{11}}$                                                                                &$\frac{\alpha_{11}^{(e)}}{2}+\frac{\alpha_{12}^{(e)}}{4}-\frac{q_{12}^2}{2 C_{11}}$                    &$\frac{1}{4}\left(\alpha_{111}+\alpha_{112}\right)$\\
O90                                 &$\frac{P_0}{\sqrt{2}}$ &$G_{44}$                           &$\alpha_1^{\rm r}+\alpha_{12}^{\rm(e)}\frac{P_0^2}{2}+\alpha_{112}\frac{P_0^4}{4}$                                                                                         &$\alpha_{11}^{(e)}+\frac{1}{2}\left[\alpha_{112}P_0^2-\frac{q_{12}^2}{C_{11}}\right]$                  &$\alpha_{111}$\\
O120                            &$\frac{\sqrt{3}P_0}{2}$&$\frac{G_{11}-G_{12}+4G_{44}}{6}$&***&***&$\frac{2\alpha_{111}+21\alpha_{112}+2\alpha_{123}}{108}$\\
\\
\hline
\multicolumn{6}{c}{Rhombohedral phase}\\
R180\{1$\bar{1}$0\}                 &$P_0$                  &$\frac{G_{11}-G_{12}+G_{44}}{3}$   &$\alpha_1+\frac{1}{6 C_{33}^\prime}[C_{12}(e_{\rm a}(q_{11} + 2 q_{12} - 4 q_{44}) - 6 e_{\rm b} q_{44})$                                                                  &$\frac{\alpha_{11}^{(e)}+\alpha_{12}^{(e)}}{3}-\frac{q_{13}^{\prime 2}}{2C_{33}^\prime}$               &$\frac{3\alpha_{111}+6\alpha_{112}+\alpha_{123}}{27}$\\
                                    &$ $                    &$ $                                &$-C_{11} (6 e_{\rm b} q_{44} + 3e_{\rm a} q_{11}^\prime)$                                                                                                                  &$ $                                                                                                    &$ $\\
                                    &$ $                    &$ $                                &$-2 C_{44} (3 e_{\rm a} (q_{11} + 2 q_{12}) + 6e_{\rm b} q_{11}^\prime)]$                                                                                                  &$ $                                                                                                    &$ $\\
R180\{$\bar{2}$11\}           &$P_0$
&$\frac{G_{11}-G_{12}+G_{44}}{3}$   &***                     &***                                   &$\frac{3\alpha_{111}+6\alpha_{112}+\alpha_{123}}{27}$\\
R109                                &$\frac{P_0}{\sqrt{3}}$ &$G_{44}$                           &$\alpha_{1}+\frac{\alpha_{12}^{(e)}P_0^2}{3}+\frac{\alpha_{112}P_0^4}{9}+\frac{2q_{12}^2 P_0^2}{3C_{11}}-\frac{q_{44}^2 P_0^2}{6C_{44}}$                                   &$\frac{\alpha_{11}^{(e)}}{2}+\frac{\alpha_{12}^{(e)}}{4}-\frac{q_{12}^2}{2C_{11}}$                                              &$\frac{\alpha_{111}+\alpha_{112}}{4}$\\
                                    &$ $                    &$ $
                    &$-e_{\rm a}(q_{11}+2q_{12})-e_{\rm b} q_{44}$          &$+\frac{1}{12}\left(2\alpha_{112}+\alpha_{123}\right)P_0^2$                                          &$ $\\
R71                                 &$\frac{P_0}{\sqrt{3}}$ &$G_{44}$                           &$\alpha_1+\frac{2\alpha_{12}^{(e)}P_0^2}{3}+\frac{2\alpha_{112}P_0^4}{9}+\frac{\alpha_{123}P_0^4}{9}$                                                                      &$\alpha_{11}^{(e)}+\frac{2\alpha_{112}P_0^2}{3}-\frac{q_{12}^2}{2C_{33}^\prime}$                       &$\alpha_{111}$\\
                                    &$ $                    &$ $                                &$-e_{\rm a}(q_{11}+2q_{12})+\frac{q_{12}^2 P_0^2}{3C_{33}^\prime}-\frac{q_{44}^2P_0^2}{3C_{44}}$                                                                           &$ $                                                                                                    &$ $\\
\hline\hline
\end{tabular}
\end{table*}

\section{\label{sec_results} QUANTITATIVE RESULTS}

\begin{table*}
\caption{Predicted values of thickness and planar energy density
of domain wall species illustrated in Fig.\,2 together with the
determining parameters appearing in the SPP treatment. Results are
evaluated from the BaTiO$_3$-specific GLD model at three selected
temperatures corresponding to the tetragonal, orthorhombic and
rhombohedral phase, respectively. Domain walls for which relaxing
of the $P_{\rm t}$ component results in a lower-energy CPP
 solution (see in Section\,VI.) are denoted by $^\dag$. Numerical
values are in SI units ($2\xi$ in nm; $\Sigma$ in mJ/m$^{2}$; $U$
in MJ/m$^3$; $p_\infty$ in C/m$^2$; $g$ in
10$^{-11}$~kg\,m$^5$\,s$^{-2}$\,C$^{-2}$; $a_1$ in
10$^7$~kg\,m$^3$\,s$^{-2}$\,C$^{-2}$; $a_{11}$ in
10$^8$~kg\,m$^7$\,s$^{-2}$\,C$^{-4}$; $a_{111}$ in
10$^9$~kg\,m$^{11}$\,s$^{-2}$\,C$^{-6}$).}\label{tab_numerical_results}
\begin{tabular}{lrrrrrrrrr}
\hline\hline
Domain wall&~~~~~~~~~~$2\xi$&~~~~~~~~~~$\Sigma$&~~~~~~~~~~~~~~$U$&~~~~~~~~~~$A$&$~~~~~~~~~~p_\infty$&~~~~~~~~~~$g$&~~~~~~~~~~$a_{1}$&~~~~~~~~~~$a_{11}$&~~~~~~~~~~$a_{111}$\\
\hline
\multicolumn{10}{c}{Tetragonal phase (298\,K)}\\
T180\{001\}                         &0.63   &5.9    &6.41   &1.43   &0.265  &2.0    &-14.26 &1.69   &8.00\\
T180\{011\}                         &0.63   &5.9    &6.41   &1.43   &0.265  &2.0    &-14.26 &1.69   &8.00\\
T90                                 &3.59   &7.0    &1.45   &1.09   &0.188  &26.5   &-7.86  &9.53   &3.12\\
\\
\hline
\multicolumn{10}{c}{Orthorhombic phase (208\,K)}\\
O180\{1$\bar{1}$0\}$^\dag$          &2.66   &31.0   &8.20   &1.70   &0.331  &26.5   &-9.71  &-2.75  &4.36\\
O180\{001\}                         &0.70   &8.9    &8.95   &1.64   &0.331  &2.0    &-11.07 &-2.13  &4.36\\
O90                                 &0.72   &4.3    &4.26   &1.50   &0.234  &2.0    &-11.62 &-0.08  &12.97\\
O60                                 &3.62   &5.3    &1.09   &1.08   &0.166  &26.1   &-7.64  &12.14  &4.36\\
O120$^\dag$                         &1.70   &13.7   &5.82   &1.47   &0.287  &10.2   &-10.82 &0.50   &4.92\\
\\
\hline
\multicolumn{10}{c}{Rhombohedral phase (118\,K)}\\
R180\{1$\bar{1}$0\}$^\dag$          &2.13   &36.0   &11.81  &1.83   &0.381  &18.3   &-9.53  &-3.64  &3.17\\
R180\{$\bar{2}$11\}$^\dag$          &2.13   &36.0   &11.81  &1.83   &0.381  &18.3   &-9.53  &-3.64  &3.17\\
R109$^\dag$                         &0.70   &7.8    &7.81   &1.65   &0.311  &2.0    &-10.84 &-2.56  &5.60\\
R71                                 &0.74   &3.7    &3.52   &1.58   &0.220  &2.0    &-10.31 &-2.42  &17.94\\
\hline\hline
\end{tabular}
\end{table*}

\begin{figure}
\centerline{
\includegraphics[width=8.6cm, clip=true]{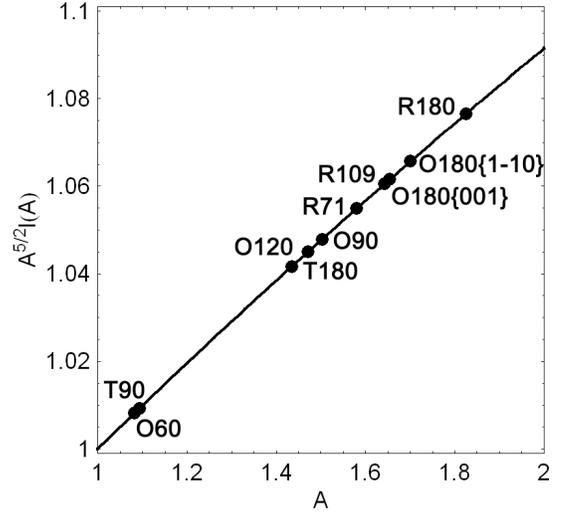}}
\caption{Dependence of correction factor in the expression for
domain wall energy density (25) as a function of the shape
coefficient $A$. Full points indicate values of $A$ for particular
domain walls considered in Table\,III.}\label{fig_integral}
\end{figure}

\begin{figure}
\centerline{
\includegraphics[width=8.6cm, clip=true]{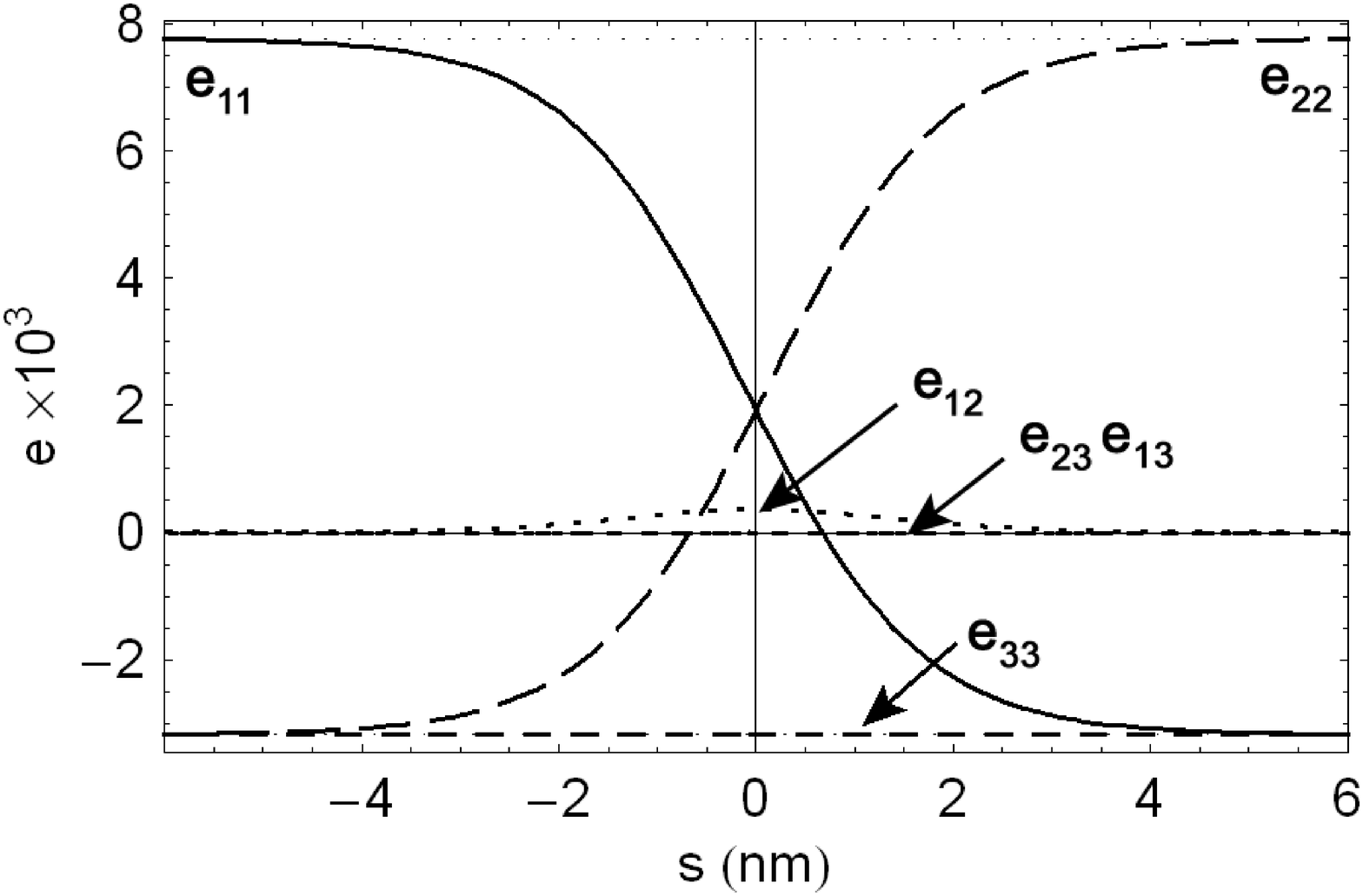}}
\caption{Course of  strain components along the domain wall normal
coordinate $s$ for the T90 domain wall. Indices refer to cubic axes
of the parent phase.}\label{fig_T90__e}
\end{figure}

\begin{figure*}
\centerline{
\includegraphics[width=17.6cm, clip=true]{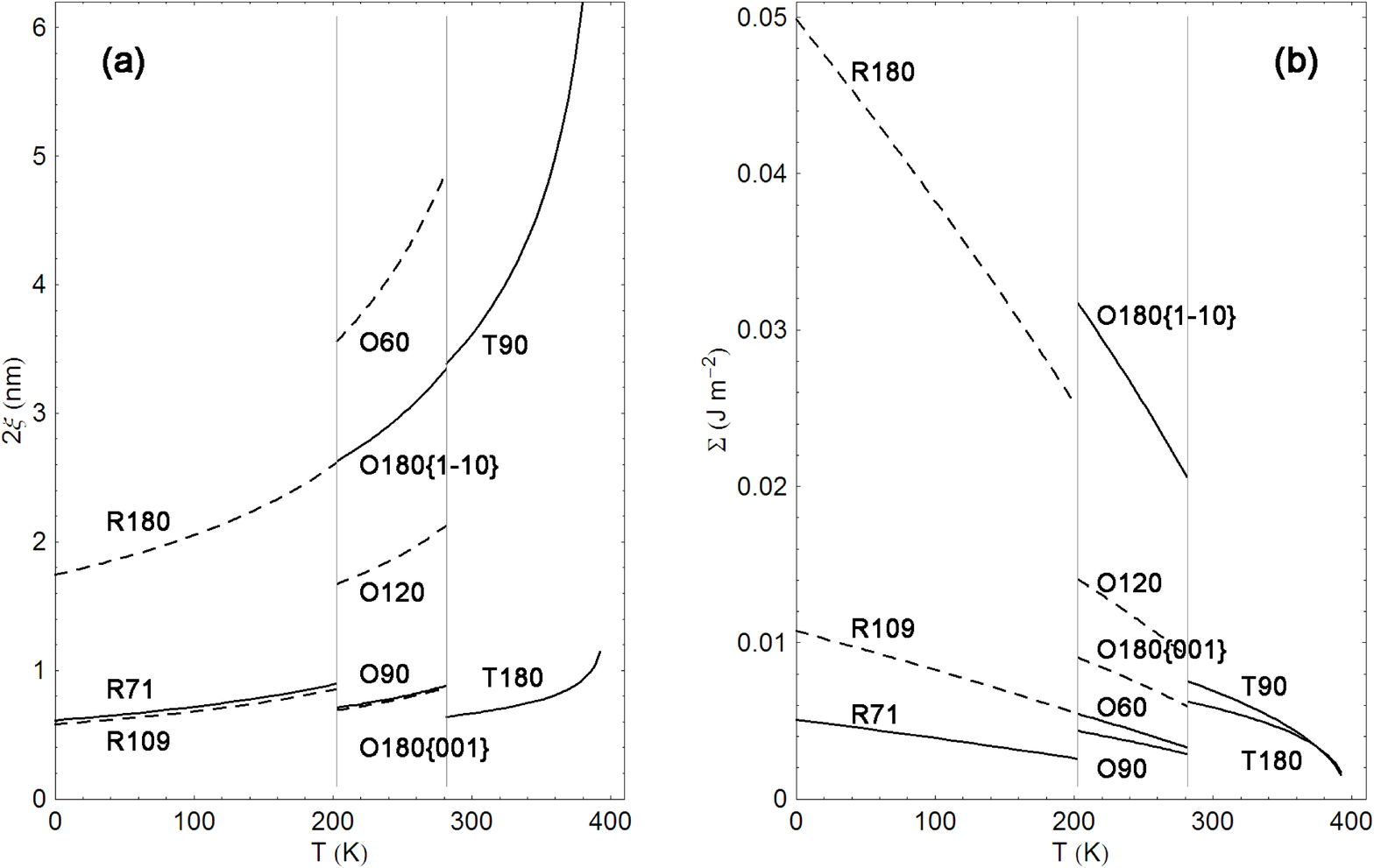}}
\caption{ Temperature dependence of the thickness (a) and energy
density (b) for various mechanically compatible and neutral domain
walls of BaTiO$_3$ estimated within SPP approximation. The SPP
values for domain walls for which relaxing of the $P_{\rm t}$
component results in a lower-energy CPP solution (see in
Section\,VI.) are shown by dashed lines. Thickness of O180\{001\},
O90 has a very close temperature dependence. The results for T180
as well as for R180 walls are not distinguished by specific
orientations of the wall normal, since for both cases the angular
dependence of thickness and energy of SPP solutions on the domain
wall normal is negligible (see Tab.\,\ref{tab_numerical_results}).
Vertical lines mark phase-transition
temperatures.}\label{fig_sirka_energie}
\end{figure*}

Advantage of the GLD approach is that domain wall properties can
be obtained at any temperature. In
Table\,\ref{tab_numerical_results} we give numerical results for
domain wall parameters at one particular temperature for each of
the ferroelectric phases: at 298\,K, 208\,K, and 118\,K for the
tetragonal, orthorhombic, and rhombohedral phase, resp.
Corresponding numerical values of the spontaneous quantities
appearing in Table\,\ref{tab_boundary_conditions} are:
$P_0=0.265\,{\rm C\,m^{-2}}$,
$e_\parallel=Q_{11}P_0^2=7.77\times10^{-3}$ and
$e_\perp=Q_{12}P_0^2=-3.18\times10^{-3}$ (the tetragonal phase,
298\,K); $P_0=0.331\,{\rm C\,m^{-2}}$, $e_{\rm
a}=\frac{Q_{11}+Q_{12}}{2}P_0^2=3.58\times10^{-3}$, $e_{\rm
c}=Q_{12}P_0^2=-4.96\times10^{-3}$ and $e_{\rm
b}=\frac{Q_{44}}{4}P_0^2=0.79\times10^{-3}$ (the orthorhombic
phase, 208\,K); $P_0=0.381\,{\rm C\,m^{-2}}$, $e_{\rm
a}=\frac{Q_{11}+2Q_{12}}{3}P_0^2=0.97\times10^{-3}$ and $e_{\rm
b}=\frac{Q_{44}}{4}P_0^2=0.70\times10^{-3}$ (the rhombohedral
phase, 118\,K). The base vectors for the O60 domain wall (see
Eqn.\,\ref{eqn_O60rst_coord}) at 208\,K are
\begin{eqnarray}
{\bf r}_{\rm O60}&=&\left(0.707,0,0.707\right)\nonumber\\
{\bf s}_{\rm O60}&=&\left(0.701,0.130,-0.701\right)\nonumber\\
{\bf t}_{\rm
O60}&=&\left(-0.092,0.991,0.092\right)~.\label{eqn_O60rstnum_coord}
\end{eqnarray}

The right four columns of the Table\,\ref{tab_numerical_results}
contain corresponding numerical values of the domain wall
coefficients $g$, $a_{1}$, $a_{11}$, and $a_{111}$ derived from
GLD parameters and spontaneous order-parameter values using above
derived analytical expressions, mostly given explicitly in
Table\,\ref{tab_analytical_expressions}. The left part of the
Table\,\ref{tab_numerical_results} contains the key domain wall
properties such as wall thickness $2\xi$ and energy density
$\Sigma$.

Clearly, T90 and O60 domain walls are considerably broader then
others. T90 wall in the tetragonal phase is predicted to be
3.59\,nm thick at 298\,K (as already calculated in
Ref.\,\onlinecite{art_hlinka_marton_prb_2006}) and S-type O60
domain wall in the orthorhombic phase is predicted to be 3.62\,nm
thick at 208\,K. Since the wall thickness is greater than the
lattice spacing, the pinning of the walls is
weak\cite{art_ishibashi} and they can be easily moved. Moreover,
in case of O60 wall, the pinning is further suppressed due to
'incommensurate' character of Miller indices of the wall normal.

As follows from Eqn.\,(\ref{eqn_analytical-wall-thickness}),
domain wall thickness is determined by quantities $g$, $U$, and
$p_{\infty}$. By inspection of their values in
Table\,\ref{tab_numerical_results}, we see that the most important
factor is the coefficient $g$. Indeed, the domain walls R71, R109,
O90, O180\{001\}, T180 with $g=2\times10^{-11}~{\rm kg\,m^5
s^{-2}\,C^{-2}}$ are all very narrow (thickness below $1$\,nm),
and the thickness of various walls monotonically increases with
increasing value of $g$. Let us stress that coefficients $a_{1}$,
$a_{11}$, $a_{111}$, and $g$ depend on the direction of the domain
wall normal. For example, O180\{1$\bar{1}$0\} with
$g=26.5\times10^{-11}~{\rm kg\,m^5 s^{-2}\,C^{-2}}$ is almost four
times broader than O180\{001\} with $g=2\times10^{-11}~{\rm
kg\,m^5 s^{-2}\,C^{-2}}$.

In the absence of other constraints, the probability of appearance of domain
wall species should be determined by the surface energy density $\Sigma$.
Therefore, in the orthorhombic phase, the thinner O180$\{001\}$ wall is more
likely to occur than the 0180$\{1\bar{1}0\}$ one. Interestingly, the
O180$\{001\}$ wall has almost the same thickness as the O90 wall, while in the
tetragonal phase, it is the 90$^\circ$ wall which is much thicker than
180$^\circ$ wall (3.59\,nm compared to 0.63\,nm).

In general, the normal of the neutral 180$^\circ$-domain wall can take any
direction perpendicular to the spontaneous polarization of the adjacent domains.
Therefore also $\Sigma$ depends on the orientation of the wall normal (in the
$s-t$ plane). We have checked in the orthorhombic phase that the 0180$\{001\}$
and 0180$\{1\bar{1}0\}$ correspond to the extremes in the angular dependence of
$\Sigma$, which is monotonous between them. This strong angular dependence is
correlated with the anisotropy of the tensor $G_{ijkl}$. However, $g$ is
independent of the wall direction in the tetragonal and rhombohedral phase, and
also the variation of coefficients $a_{1}$, $a_{11}$, and $a_{111}$ is
insignificantly small, so that the effective directional dependence of
180$^\circ$ domain wall properties in these phases is negligible.

The values of the shape factor $A$ (see Table\,\ref{tab_numerical_results})
determines the deviation of the $P_{\rm r}(s)$ polarization profile of the
domain walls from the simple $tanh$ form. For all studied cases the values of
$A$ range between 1 and 2, where the correction factor $A^{5/2}I(A)$ appearing
in the Eqn.\,(\ref{correction_factor_sigma}) is almost linear function of $A$,
as it can be seen in Fig.\,\ref{fig_integral}. It means that the shape
deviations are much smaller than those shown by broken line in
Fig.\,\ref{fig_clanek_profile}. Unfortunately, in the case of broad walls T90
and O60, which are good candidates for study of the structure of the wall
central part, $A$ is almost one.

Eqn.\,(\ref{eqn_mechanical-equilibrium-constant}) can be also used to evaluate
local strain variation in the wall. In Fig.\,\ref{fig_T90__e} the profiles of
strain tensor components for T90 domain wall are shown for illustration.
Obviously, $e_{33}, e_{23}$, and $e_{13}$ strain components are strictly
constant and equal to their boundary values as it follows from mechanical
compatibility conditions. The $e_{11}$ and $e_{22}$ components and polarization
$P_{\rm r}$ vary between their spontaneous values. Let us stress that the
're-entrant' shear component $e_{12}$ (it has the same value in both adjacent
domains) approaches the non-zero value of about 5$\times 10^{-4}$ in the middle
of the domain wall.

The temperature dependence of the thickness and surface energy density of the 12
studied domain wall species is plotted in Fig.\,\ref{fig_sirka_energie}a and
Fig.\,\ref{fig_sirka_energie}b, respectively. Although some properties do vary
considerably, e.g. thickness of T90 wall in the vicinity of the
paraelectric-ferroelectric phase transition, the sequence of the thickness
values as well as the surface energy-density values of different domain wall
types remain conserved within each phase.

The temperature dependence of domain wall thickness follows the
trend given by Eqn.\,(\ref{eqn_analytical-wall-thickness}). It
increases with increasing temperature due to the dependence on
$p_\infty$ (see dependence of $U$ on $p_\infty$ in
Eqn.\,(\ref{eqn_energy_on_pinfty})). Such behavior is well known
also from the experimental observations.\cite{art_andrews_1986,
art_robert_1996, art_huang_jiang_hu_liu_JPCM_1996}

\section{\label{sec_cpp} CURVED POLARIZATION PATH SOLUTIONS}

\begin{figure} \centerline{
\includegraphics[width=8.6cm, clip=true]{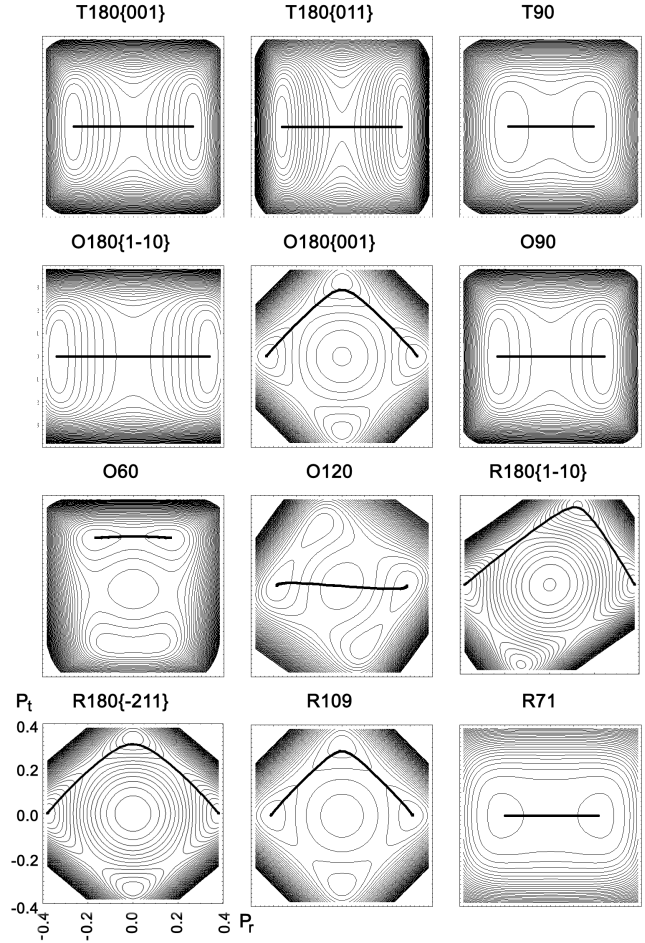}}
\caption{Equipotential contours of the Euler-Lagrange potentials
showing ELPS associated with the $P_{\rm s}$=$P_{\rm
s}(\pm\infty)$ plane for a set of BaTiO$_3$ domain walls
considered in Table\,III. Polarization scale given for the bottom
left inset (in Cm$^{-2}$) is equally valid for all shown ELPS's.
Energetically most favorable domain wall solutions with
trajectories restricted to the $P_{\rm s}$=$P_{\rm s}(\pm\infty)$
plane are indicated with bold lines. The solutions corresponding
to a linear segment are denoted in the text as SPP walls, as
opposed to the CPP solutions (see in Section\,VI.), which have a
curved order-parameter trajectory.}\label{fig_rt_pot}
\end{figure}

So far we have investigated domain walls within the SPP
approximation, i.e. components $P_{\rm s}$ and $P_{\rm t}$, which
are the same in both domain states, were kept constant inside the
whole wall. This is quite usual assumption made for a
ferroelectric domain wall. Nevertheless, the full variational
problem, where all three components of $\bf P$ could vary along
$\bf s$ coordinate, leads in general to a lower energy solution
corresponding to a curved polarization path (CPP) in the 3D
primary order-parameter space. Appearance of the non-zero
're-entrant' components within a ferroelectric domain wall was
considered e.g. in works of
Refs.\,\onlinecite{art_huang_jiang_hu_liu_JPCM_1996,
art_hlinka_marton_prb_2006, art_lee_2009,
art_meyer_vanderbilt_2002}. For 180$^\circ$ walls, the
polarization profile associated with SPP is sometimes denoted as
the Ising-type wall. In contrast, the CPP solutions with non-zero
$P_{\rm s}$, $P_{\rm t}$ are often considered as N\'eel and
Bloch-like,\cite{art_lee_2009} even though, in contrast with
magnetism, the modulus of $\bf P$ is far from being conserved
along the wall normal $\bf s$.

We have previously considered non-constant $P_{\rm s}$ component of polarization
in the 90$^\circ$ wall with explicit treatment of electrostatic interaction and
realized that the deviations from SPP approximation are quite
negligible.\cite{art_hlinka_marton_prb_2006} In general, non-constant $P_{\rm
s}$ would lead to non-vanishing $\nabla \cdot \bf {\rm P}$ and finite local
charge density, which in a perfect dielectric causes a severe energy penalty.
The same situation is expected for all domain wall species. However, there is no
such penalty for non-constant $P_{\rm t}$-solutions. It was previously
argued\cite{art_huang_jiang_hu_liu_JPCM_1996} that the Bloch-like (with a
considerable magnitude of $P_{\rm t}$ at the domain wall center) solutions could
occur in orthorhombic BaTiO$_3$. Therefore, it is interesting to systematically
check for existence of such solutions using our model.

In order to study such Bloch-like solutions, we have calculated
Euler-Lagrange potential in the order-parameter plane $P_{\rm
s}=P_{\rm s}(\pm\infty)$ by integrating Euler-Lagrange equations
(Eqn.\,\ref{eqn_mechanical-equilibrium} and
\ref{eqn_mechanical-equilibrium-constant}) for all domain wall
species from Table\,\ref{tab_numerical_results} similarly as e.g.
in the Refs.\,\onlinecite{art_cao_cross_prb_1991,
art_huang_jiang_hu_liu_JPCM_1996}. Resulting 2D Euler-Lagrange
potential surfaces (ELPS) are displayed in Fig.\,\ref{fig_rt_pot}.
In each ELPS, the bold lines indicate numerically obtained
domain-wall solution with the lowest energy. The spatial step was
chosen as 0.1\,nm, $P_{\rm r}$ and $P_{\rm t}$ were fixed to
boundary conditions in sufficient distance from domain wall
(6\,nm) and initial conditions for $P_{\rm t}$ were chosen so that
the polarization path bypasses the energy maximum of the ELPS, and
the system was relaxed to the equilibrium.

Among the twelve treated wall species, there are six cases where only the SPP
solutions with $P_{\rm t}$=const exist: T180\{001\}, T180\{011\}, T90,
O180\{1$\bar{1}$0\}, O90, and R71. These solutions are clearly "Ising-like". In
all these cases, the ($P_{\rm r}=0$, $P_{\rm t}=0$) point is the only saddle
point of the ELPS. In the other six cases -- O180\{001\}, O60, O120,
R180\{1$\bar{1}$0\}, R180\{$\bar{2}$11\}, and R109 -- the ELPS has a maximum at
the ($P_{\rm r}=0$, $P_{\rm t}=0$), and the lowest energy solutions correspond
to curved polarization paths. This suggests that the previously discussed SPP
description may not necessarily be the proper approximation for these walls.
Nevertheless, in the case of O60 and O120 walls the deviations from the SPP
model are marginal, and only the remaining four solutions exhibit strong
Bloch-like behavior. Moreover, the energy differences between SPP and CPP
solutions were found to be almost negligible, except for R180, where the CPP
energy is by about 10\% lower in the entire temperature range of stability of
the rhombohedral phase. Therefore, it is quite possible that in the case of
O180\{001\}, R180\{1$\bar{1}$0\}, R180\{$\bar{2}$11\}, and R109 walls both
Bloch-like and Ising-like solutions may be realized.

\begin{figure}
\centerline{
\includegraphics[width=8.6cm, clip=true]{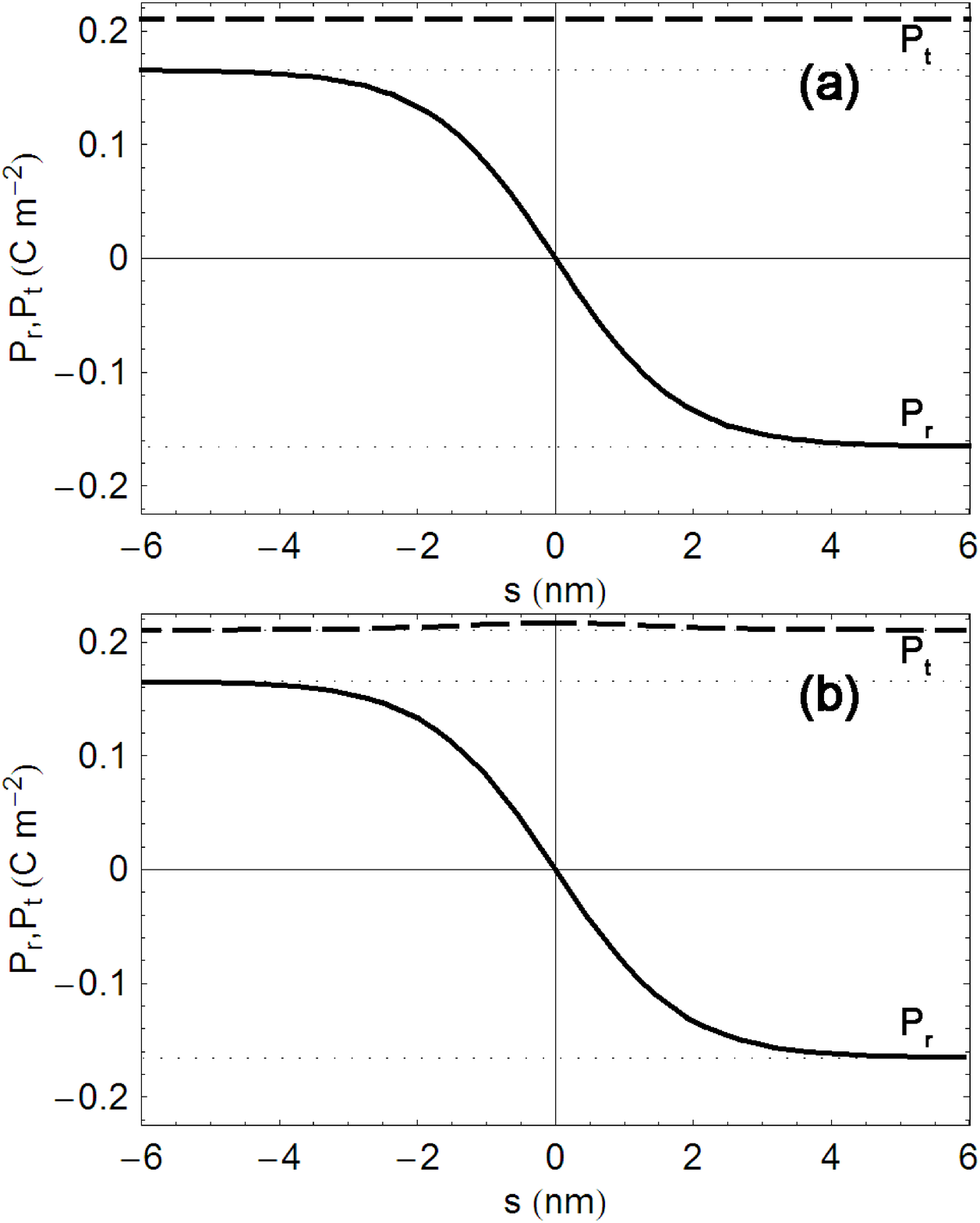}}
\caption{Predicted profiles of polarization components for O60
domain wall in BaTiO$_3$ at $T= 208\,$K. Full line stands for the
$P_{\rm r}$ component of polarization vector, broken line for the
$P_{\rm t}$ one. Analytically obtained SPP solution (a) practically
does not differ from the numerically obtained CPP
 solution (b) with an unconstrained $P_{\rm t}$ component in
this case.}\label{fig_p_profiles_os}
\end{figure}

The deviations from SPP in the case of the almost Ising-like O60 and O120 walls
are associated with the fact that ELPS is not symmetric with respect to $P_{\rm
t}=P_{\rm t}(\pm\infty)$ mirror plane. In these cases not only the polarization
path and wall energies, but also domain wall thicknesses of the SPP and CPP
counterpart solutions are very similar. This is demonstrated in
Fig.\,\ref{fig_p_profiles_os}, which shows polarization profiles of both SPP and
CPP solutions for the O60 wall.

\begin{figure}
\centerline{
\includegraphics[width=8.6cm, clip=true]{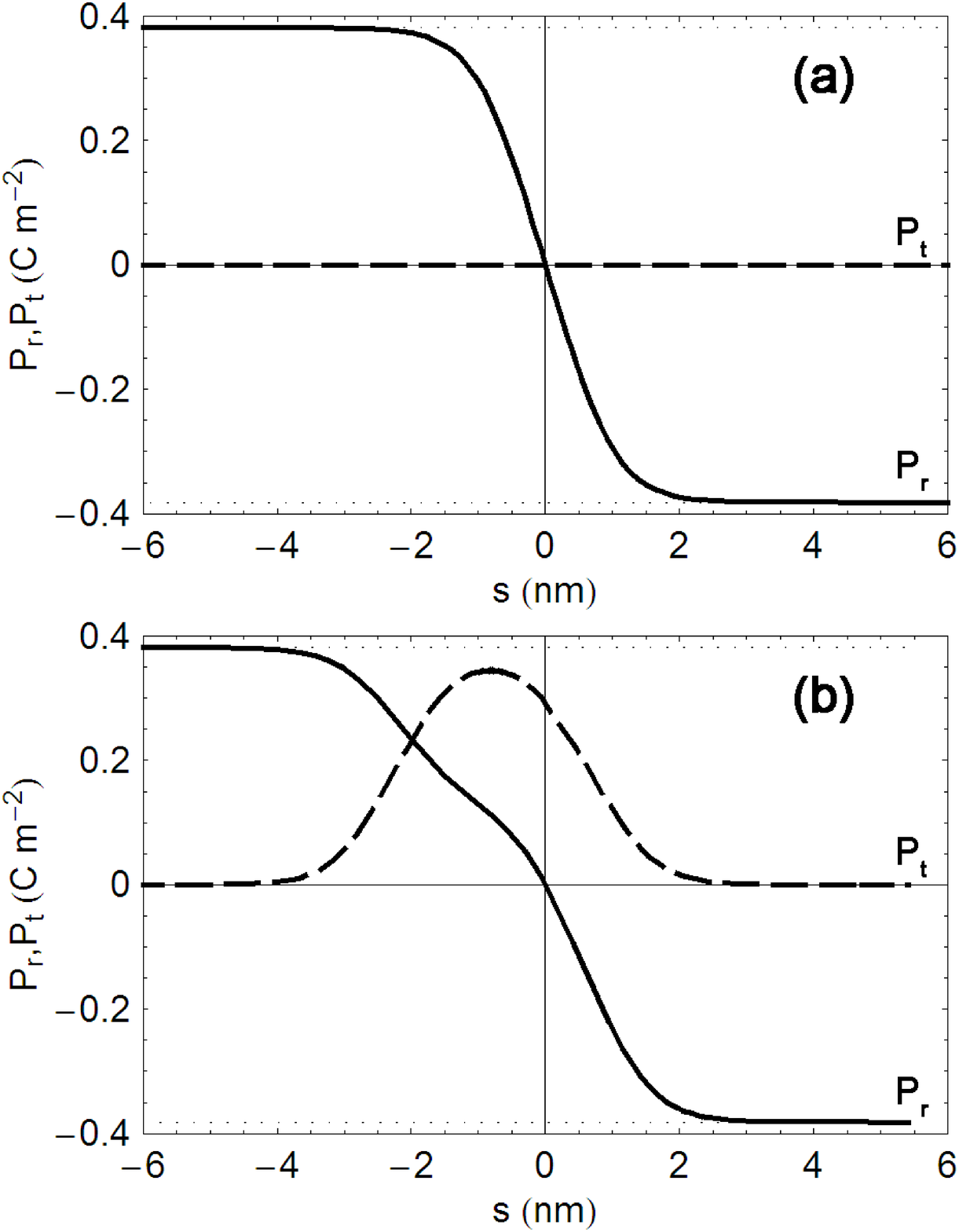}}
\caption{Profiles of polarization components in R180\{1$\bar{1}$0\}
domain wall showing the SPP stationary trajectory (a) as well as the
lower energy Bloch-like solution (b), both for the model parameters
corresponding to $T= 118\,$K. Full line stands for the $P_{\rm r}$
component of polarization vector, broken line for the $P_{\rm t}$
one. The bottom panel demonstrates that Bloch solution may
considerable modify the domain domain wall
profile.}\label{fig_p_profiles_r180}
\end{figure}

Much more pronounced difference between domain wall profiles of SPP and CPP
solutions are found in case of O180\{001\}, R180\{1$\bar{1}$0\},
R180\{$\bar{2}$11\}, and R109 Bloch-like walls where the CPP trajectories bypass
the ($P_{\rm r}=0$, $P_{\rm t}=0$) maximum near the additional minima, which
originate from 'intermediate' domain states either of the same phase or even of
the different ferroelectric phases.  In these cases, the inadequacy of SPP
approximation is obvious. For example, the CPP of R180\{$\bar{2}$11\} wall seems
to pass through a additional minimum corresponding to an 'orthorhombic'
polarization state (consult corresponding inset in the Fig.\,\ref{fig_plochy}).
As expected, the profile of such CPP solution deviates strongly from $tanh$
shape and even definition of the wall thickness would be problematic (see
Fig.\,\ref{fig_p_profiles_r180}).

\section{\label{sec_conclusion}CONCLUSION}

The work reports detailed study of mechanically compatible and electrically
neutral domain walls in BaTiO$_3$. The investigation was done within the
framework of the GLD model. Using the SPP approximation it was possible to
compare properties of various kinds of domain wall species from the same
perspective.

The phenomenological nature of the GLD model allowed to predict the temperature
dependence of the domain wall characteristics in the whole temperature range of
ferroelectric phases. Its continuous nature gave us even the opportunity to
deal conveniently with the non-crystallographic S-type domain wall, which has a
general orientation with respect to the crystal lattice, and which is therefore
difficult to cope with in discrete models relying on periodic boundary
conditions.

The S-wall in the orthorhombic, as well as the 90$^\circ$ wall in tetragonal
phase were both found to be about 4\,nm thick and consequently are expected to
be mobile, i.e. they could be easily driven by external fields, and they may
thus significantly contribute to the dielectric or piezoelectric response of the
material.

For several temperatures, we have numerically investigated domain walls allowing
for more complicated CPP solutions with non-constant $P_{\rm t}$. We have
identified solutions, which could be considered as analogues of Bloch walls
known from magnetism. Interestingly, in contrast with
Ref.\,\onlinecite{art_huang_jiang_hu_liu_JPCM_1996}, our model predicts the
Ising-type profile of the O180\{1$\bar{1}$0\} wall. At the same time the
Bloch-like structure of the O180\{001\} wall is predicted.

We believe that this kind of somewhat exotic walls  actually represent important
generic examples of ferroelectric domain species, which should be anticipated in
all ferroelectrics with several equivalent domain states distinguished
simultaneously by the orientation of the spontaneous polarization and strain.
They should be certainly taken into account in investigations of domain-wall
phenomena in ferroelectric perovskites. At the same time, the energy differences
between the Bloch-like and Ising-like solutions are rather subtle here.
Therefore, in spite of the fairly good agreement for 180$^\circ$ and 90$^\circ$
domain walls between ab-initio calculations and predictions of this model in the
tetragonal phase,\cite{art_hlinka_marton_prb_2006} the preference for the
calculated Bloch-like trajectories may not necessarily reproduced for the domain
walls encountered in real BaTiO$_3$ crystal, since there is obviously a
considerable uncertainty in the adopted material-specific GLD parameters. In
addition,  predictions for the domain walls with very small thickness must be
considered with a particular caution since the description  of the sharp domain
wall profiles obviously touches the limits of the applicability of the
continuous model.

In conclusion, we have derived a number of qualitative and quantitative
predictions for mechanically compatible neutral domain walls of tetragonal,
orthorhombic and rhombohedral BaTiO$_3$ on the basis of the previously proposed
material-specific GLD model. We believe that the insight into the domain wall
properties mediated by the provided analytical and numerical analysis could be
helpful for understanding of domain wall phenomena in BaTiO$_3$ as well as in
some other intensively investigated members of the ferroelectric perovskite
family with same sort of macroscopic ferroelectric phases, for example in
KNbO$_3$, BiFeO$_3$, PbTiO$_3$ or even PZT and perovskite relaxor-related
materials.

\begin{acknowledgments}
Authors are grateful to Prof. V. Janovec for helpful comments and
critical reading of the manuscript. The work has been supported by
the Czech Science Foundation (Projects Nos. P204/10/0616 and
202/09/0430).
\end{acknowledgments}

\end{document}